\documentclass{article}
\usepackage[margin=1in]{geometry}
\usepackage{amsmath}
\usepackage{epsfig}
\numberwithin{equation}{section} % amsmath

\newcommand{\beq}{\begin{equation}}
\newcommand{\eeq}{\end{equation}}
\newcommand{\beqa}{\begin{eqnarray}}
\newcommand{\eeqa}{\end{eqnarray}}
\newcommand{\bdm}{\begin{displaymath}}
\newcommand{\edm}{\end{displaymath}}
\newcommand{\lslash}[1]{#1\llap/}

\newcommand{\Eq}[1]{Eq.\ (\ref{#1})}
\newcommand{\Eqs}[2]{Eqs.\ (\ref{#1}) and (\ref{#2})}

\newcommand{\Rref}[1]{Ref.\ \cite{#1}}
\newcommand{\Rrefs}[2]{Refs.\ \cite{#1} and \cite{#2}}

\newcommand{\Fig}[1]{Fig.\ \ref{#1}}

\newcommand{\Figss}[3]{Figs.\ \ref{#1}, \ref{#2} and \ref{#3}}
\newcommand{\Table}[1]{Table\ \ref{#1}}
\newcommand{\Section}[1]{Section\ \ref{#1}}

\newcommand{\vkappa}{\vec\kappa}
\newcommand{\omeganu}{\omega^{(\nu)}}
\newcommand{\omeganub}{\omega^{(\bar\nu)}}
\title{
  Neutrino effective potential and damping in a fermion and scalar
  background in the resonance region
}

\author{Jos\'e F. Nieves\footnote{nieves@ltp.uprrp.edu}\\
  Laboratory of Theoretical Physics, Department of Physics\\
  University of Puerto Rico, R\'{\i}o Piedras, Puerto Rico 00936
  \and\\[12pt]
  Sarira Sahu\footnote{sarira@nucleares.unam.mx}\\
  Instituto de Ciencias Nucleares\\
  Universidad Nacional Aut\'onoma de Mexico\\
  Circuito Exterior, C. U.\\
  A. Postal 70-543, 04510 Mexico DF, Mexico\\
}

\date{January 2022}

\begin{document}
\maketitle

\begin{abstract}
  We consider the propagation of a neutrino or an
  antineutrino in a medium composed of fermions ($f$) and scalars ($\phi$)
  interacting via a Yukawa-type coupling of the form $\bar f\nu\phi$, for
  neutrino energies at which the processes like $\nu + \phi \leftrightarrow f$
  or $\nu + \bar f \leftrightarrow \bar\phi$, and the corresponding ones for
  the antineutrino, are kinematically accessible. The relevant energy values
  are around $|m^2_\phi - m^2_f|/2m_\phi$ or $|m^2_\phi - m^2_f|/2m_f$,
  where $m_\phi$ and $m_f$ are the masses of $\phi$ and $f$, respectively.
  We refer to either one of these regions as a \emph{resonance energy range}.
  Near these points, the one-loop formula for the neutrino self-energy
  has a singularity. From a technical point of view, that feature
  is indicative that the self-energy acquires an imaginary part, which is
  associated with damping effects and cannot be neglected, while
  the integral formula for the real part must be
  evaluated using the principal value of the integral.
  We carry out the calculations explicitly for some cases
  that allow us to give analytic results.
  Writing the dispersion relation in the form
  $\omega = \kappa + V_{\text{eff}} - i\gamma/2$, we give the explicit formulas
  for $V_{\text{eff}}$ and $\gamma$ for the cases considered.
  When the neutrino energy is either much larger or much smaller
  than the resonance energy, $V_{\text{eff}}$ reduces to the
  effective potential that has been already determined in the literature
  in the high or low momentum regime, respectively. 
  The virtue of the formula we give for $V_{\text{eff}}$ is that it is
  valid also in the \emph{resonance energy range}, which is outside
  the two limits mentioned.
  As a guide to possible applications we give the relevant formulas
  for $V_{\text{eff}}$ and $\gamma$, and consider the solution to the
  oscillation equations including the damping term, in
  a simple two-generation case.
\end{abstract}

\section{Introduction and Summary}
\label{sec:introduction}

As is well known, the properties of neutrinos
that propagate through a medium differ from those in the
vacuum. In particular, the energy-momentum for massless neutrinos
$\omega = \kappa$, where $\omega$ is the energy and $\kappa$ the
magnitude of the momentum vector, is not valid in the
medium\cite{wolfenstein,langacker}. The modifications of the neutrino dispersion
relation can be represented in terms of an index of refraction, or
more suitable for our purposes, in terms of an effective
potential ($V_{\text{eff}}$) and a damping ($\gamma$), by writing
it in the generic form
\beq
\label{drgeneric}
\omega = \kappa + V_{\text{eff}} - i\gamma/2\,.
\eeq
It is now well established that an efficient method to determine
the dispersion relation, is to compute $V_{\text{eff}}$ and $\gamma$ from the
calculation of the neutrino thermal
self-energy\cite{weldon-fermions,notzold,palpham,nu-inmedium}
in the framework of thermal field theory\cite{ftft:reviews}.

In several models and extensions of the standard electro-weak theory
the neutrinos interact with scalar particles ($\phi$)
and fermions ($f$) via a coupling of the form $\bar f_R\nu_{L}\phi$ or
just with neutrinos $\bar\nu^c_R\nu_{L}\phi$.
Couplings of the latter form have been explored recently in various
contexts\cite{nunuphi,Farzan:2018gtr,%
Stephenson:1993rx,hm:2017mio,Heurtier:2016otg,%
Sawyer:2006ju,Pasquini:2015fjv,shao-feng,Brdar:2017kb}.
Such couplings produce additional contributions
beyond the standard ones to the neutrino effective
potential when the neutrino propagates in a neutrino background,
as it occurs in the environment of a supernova, where
the effect leads to the collective neutrino oscillations
and related phenomena(see for example \Rrefs{Duan:2010bg}{Chakraborty:2016yeg}
and the works cited therein),
or in the hot plasma of the Early-Universe before the neutrinos
decouple\cite{Wong:2002fa,Mangano:2006ar}.
Couplings of the form $\bar f_R\nu_{L}\phi$
produce additional contributions to the neutrino effective potential
when the neutrino propagates in a background of $\phi$ and $f$ particles
and their possible effects have been considered in
the context of Dark Matter-neutrino interactions\cite{Mangano:2006mp,
  Binder:2016pnr,
  Primulando:2017kxf,Campo:2017nwh,Brune:2018sab,Franarin:2018gfk,Dev:2019anc,
  Pandey:2018wvh,Karmakar:2018fno}.
More recently it has been pointed out that observable effects of such
scalar interactions, although precluded in terrestrial experiments,
are still possible in future solar and supernovae neutrino data,
and in cosmological observations such as cosmic microwave background
and big bang nucleosynthesis data\cite{babu}.

Motivated by these developments, we carried out in previous work
a systematic calculation of the neutrino effective potential
in such models\cite{nsnuphireal}. We considered various cases,
depending on the magnitude of $\kappa$ relative to other
parameters such as the masses of the particles and the temperature
of the background. In the limit of small $\kappa$, the effective
potential becomes independent of $\kappa$ and has a form
that is reminiscent of the Wolfenstein potential\cite{wolfenstein}.
In the opposite limit (relatively large $\kappa$), the effective
potential has a term proportional to $1/\kappa$ that mimics
a contribution to the neutrino mass\cite{ge-parke}.

The main point that is relevant to the present work, is that
in the intermediate $\kappa$ region, to be defined precisely below,
neither limiting case is a valid approximation to the effective
potential. From a physical point of view, there is a region
of neutrino energies at which the processes like $\nu + \phi \leftrightarrow f$
or $\nu + \bar f \leftrightarrow \bar\phi$, and the corresponding ones for
the antineutrino, become kinematically accessible. The relevant energy values
are around $|m^2_\phi - m^2_f|/2m_\phi$ or $|m^2_\phi - m^2_f|/2m_f$,
to which we refer as a \emph{resonance energy range}.
In those ranges, the one-loop integral formula for the neutrino self-energy
has a singularity, as has been emphasized recently in \Rref{smirnov}.

From a technical point of view, the singularity
is indicative of two things. Firstly, at those points the self-energy
acquires an imaginary part that cannot be neglected.
The imaginary part of the self-energy is associated with damping effects,
and determines the damping term $\gamma$ in the dispersion relation.
A systematic calculation of the damping terms was carried out in
\Rref{nsnuphidamp}.

Secondly, and what is our main observation here,
the effective potential, which is determined from the real (dispersive)
part of the self-energy, must be evaluated using the principal value of
the integral formula for the self-energy. The principal value
prescription allows us to give a well-defined meaning to
(the real part of) the integral for values of $\kappa$ around
the singularities. Our purpose in this work is to carry out the calculation
of the effective potential in the resonance regions
using the strategy just explained.
Writing the dispersion relation in the form given in \Eq{drgeneric},
we give the explicit formulas for $V_{\text{eff}}$ and $\gamma$
for some cases that allow us to give analytic results, and indications for
carrying out extensions and generalizations to other cases of interest.
When the neutrino energy is either much larger or much smaller
than the resonance energy, $V_{\text{eff}}$ reduces to the
effective potential that has been already determined in the literature
in the high or low momentum regime, respectively. 
The virtue of the formula we give for $V_{\text{eff}}$ is that it is
valid also in the \emph{resonance energy range}, which is outside
the two limits mentioned. As a guide to possible applications to neutrino
oscillations in the case that the neutrino energy is in the resonance region,
we give the relevant formulas for the $V_{\text{eff}}$ and $\gamma$ terms
that enter in the oscillation equations in a simple two-generation case,
and consider their solution including the damping term.

In Section \ref{sec:preliminaries} we summarize our notation and conventions,
and the context in which we carry out the calculations.
In Section \ref{sec:resonance-neutrino} we calculate the effective
potential, paying special attention to the contribution from the resonance
terms, which are expressed as an integral over the background particle
momentum distribution functions. We consider in detail
the evaluation of the relevant integral for the case that the
resonance term is the fermion background contribution,
and for concreteness we give the explicit formulas for the
case of a non-relativistic and degenerate Fermi gas.
Such formulas can be particularly useful for considering
the implications and/or setting limits on the
neutrino interactions with light particle dark matter backgrounds from
their effect on the phenomenology in reactor, solar, atmospheric,
and accelerator experiments.
In Section \ref{sec:damping-Omegaf} we consider the damping term.
We discuss some generalizations and possible extensions of the results in
Section \ref{subsec:resonance-antineutrino-Omegaf},
and the two-generation example case mentioned above
in Section \ref{subsec:twogenexample}.

\section{Preliminaries}
\label{sec:preliminaries}

In this section we review the context of the present work
and state the problem on which we focus.

\subsection{Context}

For definiteness we consider only one neutrino flavor coupling
to the fermion and scalar, which we denote simply by $\nu$,
and write
\beq
\label{Lintone}
L^{(f\nu\phi)} = \lambda\bar f_R\nu_{L}\phi + h.c\,.
\eeq
We denote by $k^\mu$ the momentum four-vector
of the propagating neutrino and by $u^\mu$ the velocity
four-vector of the background medium. In the background medium's own
rest frame, $u^\mu$ takes the form
\beq
\label{restframe}
u^\mu = (1,\vec 0)\,,
\eeq
and in this frame we write
\beq
\label{krestframe}
k^\mu = (\omega,\vec\kappa)\,.
\eeq
Since we are considering only one background medium, we can take it
to be at rest and therefore we adopt \Eqs{restframe}{krestframe}
throughout. For completeness, we briefly review and borrow
from \Rrefs{nsnuphireal}{nsnuphidamp} the
formulas that we will use to determine the dispersion relations
from the self-energy calculation. We remind that the calculations
are based on the application of the real-time Thermal Field Theory methods.

The neutrino dispersion relation is determined by the solution
of the equation
\beq
\label{eqmotion}
(\lslash{k} - \Sigma_{\text{eff}})\psi_L(k) = 0\,,
\eeq
where $\Sigma_{\text{eff}}$ is the neutrino thermal self-energy.
$\Sigma_{\text{eff}}$ can be decomposed into its dispersive ($\Sigma_r$)
and absorptive ($\Sigma_i$) parts,
\beq
\label{Sigmaeffdecomp}
\Sigma_{\text{eff}} = \Sigma_r + i\Sigma_i\,.
\eeq
$\Sigma_r$ is given by the real (dispersive) part of the
11 element of the neutrino thermal self-energy matrix,
while $\Sigma_i$ is determined from the 12 element.

The chirality of the neutrino interactions implies that $\Sigma_{\text{eff}}$
has the form
\beq
\Sigma_{\text{eff}} = \lslash{V}L\,,
\eeq
where $L = \frac{1}{2}(1 - \gamma_5)$. Corresponding to the
decomposition of $\Sigma_{\text{eff}}$ in \Eq{Sigmaeffdecomp},
\beq
\label{Vdecomp}
V^\mu = V^\mu_r + iV^\mu_i\,,
\eeq
where $V^\mu_r$ and $V^\mu_i$ are real. $V_r$ and $V_i$ are functions
of $\omega$ and $\vec\kappa$, but to simplify the notation we omit
writing the arguments unless it is necessary to indicate them.

Equation (\ref{eqmotion}) has two solutions. Denoting them by
$\omega^{(\lambda)}$ ($\lambda = \pm$), they are determined by the equation
\beq
\omega^{(\lambda)} = V^0(\omega^{(\lambda)},\vkappa) +
      \lambda [(\vkappa - \vec V(\omega^{(\lambda)},\vkappa))
        \cdot(\vkappa - \vec V(\omega^{(\lambda)},\vkappa))]^{1/2}\,,
\eeq
or to lowest order,
\beq
\label{drbase}
\omega^{(\lambda)} = V^0(\omega^{(\lambda)},\vkappa) + \lambda(\kappa -
\hat\kappa\cdot\vec V(\omega^{(\lambda)},\vkappa))\,.
\eeq
The neutrino ($\omeganu$) and antineutrino ($\omeganub$)dispersion
relations are identified as
\beqa
\omeganu(\vkappa) & = & \omega^{(+)}(\vkappa)\,,\nonumber\\
\omeganub(\vkappa) & = & \left(-\omega^{(-)}(-\vkappa)\right)^\ast\,.
\eeqa
Decomposing $\omeganu$ and $\omeganub$ in terms of their real
and imaginary parts in the form ($x = \nu,\bar\nu$),
\beq
\label{defgammanu}
\omega^{(x)} = \omega^{(x)}_r - \frac{i}{2}\gamma^{(x)}\,,
\eeq
and assuming that it is a valid approximation to set
\beqa
\label{stdapproximation}
\gamma^{(x)}_i & \ll & |\omega^{(x)}_r|\,,\nonumber\\
\omega^{(x)}_r & \simeq & \kappa\,,
\eeqa
\Eq{drbase} gives, for the real part
\beqa
\label{nudr1}
\omeganu_r & = & \kappa + V^{(\nu)}_{\text{eff}}(\vkappa)\,,\nonumber\\
\omeganub_r & = & \kappa + V^{(\bar\nu)}_{\text{eff}}(\vkappa)\,,
\eeqa
while for the imaginary part
\beqa
\label{gammastandard}
-\frac{1}{2}\gamma^{(\nu)} & = & 
\frac{n\cdot V_i(\kappa,\vec\kappa)}
{1 - n\cdot\left.
\frac{\partial V_r(\omega,\vec\kappa)}{\partial\omega}
\right|_{\omega = \kappa}}\,,\nonumber\\
-\frac{1}{2}\gamma^{(\bar\nu)} & = & 
\frac{n\cdot V_i(-\kappa,-\vec\kappa)}
{1 - n\cdot\left.
\frac{\partial V_r(\omega,-\vec\kappa)}{\partial\omega}
\right|_{\omega = -\kappa}}\,,
\eeqa
where
\beqa
\label{Veffstandard}
V^{(\nu)}_{\text{eff}}(\vkappa) & = & n\cdot V_r(\kappa,\vkappa)\,,\nonumber\\
V^{(\bar\nu)}_{\text{eff}}(\vkappa) & = & -n\cdot V_r(-\kappa,-\vkappa)\,,
\eeqa
with
\beq
n^\mu = (1,\hat\kappa)\,.
\eeq
In those cases in which the
correction due to the $n\cdot\partial V_r(\omega,\vec\kappa)/\partial\omega$
in the denominator can be neglected, the formulas in \Eq{gammastandard}
simplify to
\beqa
\label{nudisprelimg-simple}
-\frac{\gamma^{(\nu)}(\vec\kappa)}{2} & = & 
n\cdot V_i(\kappa,\vec\kappa)\,,\nonumber\\
-\frac{\gamma^{(\bar\nu)}(\vec\kappa)}{2} & = & 
n\cdot V_i(-\kappa,-\vec\kappa)\,,
\eeqa
which are the ones we will use here, borrowing from the work
in \Rref{nsnuphidamp}.

\subsection{Statement of the problem}

To state the problem in concrete terms and set the stage for the
work that follows, we recall (see, e.g, \Rref{nsnuphireal})
the following expression for the background-dependent part of the
11 element of the thermal self-energy matrix
in the $f$ and $\phi$ background,
\beq
\label{Sigma11}
\Sigma_{11}(k) = \Sigma^{(f)}_{11}(k) + \Sigma^{(\phi)}_{11}(k)\,,
\eeq
where\footnote{%
  We take the opportunity to mention that by an abuse in notation
  the symbols $\eta_F(p,\alpha_f)$ and $\eta_B(p,\alpha_\phi)$
  used in Eqs (20) and (21) in \Rref{nsnuphireal} are the same as the
  $\eta_f(p)$ and $\eta_\phi(p)$ defined in Eq (17) of that reference,
  and reproduced below in \Eq{etafphi}.
  }
\beqa
\label{Sigmaf11}
\Sigma^{(f)}_{11} & = & -|\lambda|^2\int\frac{d^4p}{(2\pi)^3}
\frac{\lslash{p}L}{(p - k)^2 - m^2_\phi + i\epsilon}
\delta(p^2 - m^2_f)\eta_f(p)\,,\\
\label{Sigmaphi11}
\Sigma^{(\phi)}_{11} & = & |\lambda|^2\int\frac{d^4p}{(2\pi)^3}
\frac{(\lslash{p} + \lslash{k})L}{(p + k)^2 - m^2_f + i\epsilon}
\delta(p^2 - m^2_\phi)\eta_\phi(p)\,.
\eeqa
Using the label $x$ to stand for either $f$ or $\phi$,
the functions $\eta_x(p)$ are given by
\beq
\label{etafphi}
\eta_x(p) = \theta(p^0)f_x(p^0) +
\theta(-p^0)f_{\bar x}(-p^0)\,,
\eeq
where  $f_{x,\bar x}(p^0)$ are the equilibrium
momentum distribution functions of the background particles and antiparticles,
\beqa
\label{thermaldist}
f_{f,\bar f}(p^0) & = &
\frac{1}{e^{\beta p^0 \mp \alpha_f} + 1}\,,\nonumber\\
f_{\phi,\bar\phi}(p^0) & = & 
\frac{1}{e^{\beta p^0 \mp \alpha_\phi} -1}\,,
\eeqa
where $\beta = 1/T$ and $\alpha_x = \beta\mu_x$, with
$T$ being the temperature and $\mu_x$ the chemical potentials.

To be precise, we mention that in \Eq{Sigma11} we are
neglecting the term that involves the
product of the two thermal parts of the propagators, which does
not contribute to the real part of $\Sigma_{11}$. Thus,
going back to \Eq{Sigmaeffdecomp}, the dispersive part $\Sigma_r$
is given by
\beq
\Sigma_r = \Sigma^{(f)}_r + \Sigma^{(\phi)}_r\,,
\eeq
where
\beqa
\label{Sigmafr}
\Sigma^{(f)}_{r} & = & -|\lambda|^2 \int\frac{d^4p}{(2\pi)^3}
\frac{\lslash{p}L}{(p - k)^2 - m^2_\phi}
\delta(p^2 - m^2_f)\eta_f(p)\,,\\
\label{Sigmaphir}
\Sigma^{(\phi)}_{r} & = & |\lambda|^2 \int\frac{d^4p}{(2\pi)^3}
\frac{(\lslash{p} + \lslash{k})L}{(p + k)^2 - m^2_f}
\delta(p^2 - m^2_\phi)\eta_\phi(p)\,.
\eeqa
In \Eqs{Sigmafr}{Sigmaphir}, and the integrals that follow, the integrations
are to be interpreted in the sense of their principal value.

Carrying out the integral over $p^0$, we obtain
\beqa
\label{Vrv}
n\cdot V_r(\omega,\vkappa) & = & v_f(\omega,\vkappa) +
v_{\bar f}(\omega,\vkappa) + v_\phi(\omega,\vkappa) +
v_{\bar\phi}(\omega,\vkappa)\,,
\eeqa
where
\begin{subequations}
\label{vterms}
\begin{align}
\label{vf}
v_f(\omega,\vkappa) & =
-|\lambda|^2 \int\frac{d^3p}{(2\pi)^3 2E_f}(n\cdot p)
\frac{f_f(\vec p)}{D_f(k,p)}\,,\\
\label{vfbar}
v_{\bar f}(\omega,\vkappa) & = 
|\lambda|^2 \int\frac{d^3p}{(2\pi)^3 2E_f}(n\cdot p)
\frac{f_{\bar f}(\vec p)}{D_f(k,-p)}\,,\\
\label{vphi}
v_\phi(\omega,\vkappa) & =
|\lambda|^2 \int\frac{d^3p}{(2\pi)^3 2E_\phi}
\frac{(n\cdot p + (\omega - \kappa))f_\phi(\vec p)}{D_\phi(k,p)}\,,\\
\label{vphibar}
v_{\bar\phi}(\omega,\vkappa) & =
|\lambda|^2 \int\frac{d^3p}{(2\pi)^3 2E_\phi}
\frac{(-n\cdot p + (\omega - \kappa))f_{\bar\phi}(\vec p)}{D_\phi(k,-p)}\,,
\end{align}
\end{subequations}
with
\beq
E_x = \sqrt{{\vec p\,}^2 + m^2_x}\,, \qquad (x = f,\phi)\,,
\eeq
\beqa
D_f(k,p) & = & (p - k)^2 - m^2_\phi = k^2 - 2p\cdot k + m^2_f - m^2_\phi\,,
\nonumber\\
D_\phi(k,p) & = & (p + k)^2 - m^2_f = k^2  + 2p\cdot k + m^2_\phi - m^2_f\,,
\eeqa
and
\beq
\label{ncdotp}
n\cdot p = E_x(1 - \hat\kappa\cdot\vec v_x)\,.
\eeq
In \Eq{ncdotp} $\vec v_x$ stands for the velocity of the background particle.

To bring out the issue that we want to address,
consider for example the contribution from the $\bar f$ background,
and suppose that the conditions are such that it can be treated
in the non-relativistic limit. Then approximating
\beq
\label{pzero}
p^\mu \rightarrow (m_f,\vec 0)\,,
\eeq
in the integrand, $v_{\bar f}(\kappa,\vkappa)$
is inversely proportional to 
\beq
\label{highlowkappa}
\left.D_f(k,-p)\right|_{\omega = \kappa} \simeq 2m_f\omega -
(m^2_\phi - m^2_f)\,.
\eeq
Identifying the effective potential by \Eq{Veffstandard},
in the low momentum (\emph{heavy background}) limit this gives
a momentum-independent contribution to the effective potential
reminiscent of the standard Wolfenstein term. In the opposite limit,
the high momentum (or \emph{light background}) limit this gives
a term proportional to $1/\kappa$ that mimics
a contribution to the neutrino mass\cite{ge-parke}.
But in the intermediate region the expression is not valid and
actually undefined at
\beq
\omega \sim \frac{m^2_\phi - m^2_f}{2m_f}\,.
\eeq
If $m_\phi > m_f$, physically this feature reflects the fact that
in that regime the process $\nu + \bar f \rightarrow \bar\phi$ is kinematically
accessible. In some cases the singularity appears for $\omega$ negative,
which corresponds to the antineutrino dispersion relation. This is the
case, in the example above, if $m_f > m_\phi$, and in that case
the singularity corresponds to the process
$\bar f \rightarrow \bar \nu + \bar\phi$. Similar considerations
apply to the $v_f$ and  the $\phi$ background terms. An exhaustive
list of the various possibilities is summarized in \Table{table:cases}.
\begin{table}
  \begin{center}
    \begin{tabular}{|c|c|c|c|c|}
      \hline
      {} & $v_f$ & $v_{\bar f}$ & $v_\phi$ & $v_{\bar\phi}$\\
      \hline
      $m_\phi > m_f$ &
        $-\Omega_f $ & $\Omega_f$ & $-\Omega_\phi$ & $\Omega_\phi$\\
        {} &
        $\bar\nu + f\rightarrow\phi$ & $\nu + \bar f\rightarrow\bar\phi$ &
        $\phi\rightarrow\bar\nu + f$ & $\bar\phi\rightarrow\nu + \bar f$\\
      \hline
      $m_f > m_\phi$ &
        $\Omega_f$ & $-\Omega_f$ & $\Omega_\phi$ & $-\Omega_\phi$\\
        {} &
        $f\rightarrow\nu + \phi$ & $\bar f\rightarrow\bar\nu + \bar\phi$ &
        $\nu + \phi\rightarrow f$ & $\bar\nu + \bar\phi\rightarrow \bar f$\\
      \hline
  \end{tabular}
  \caption{
    \label{table:cases}
    List of possible resonance conditions in each background contribution and
    the associated physical processes. The parameters $\Omega_{x}$ are
    defined as $\Omega_x = |m^2_\phi - m^2_f|/2m_x$ (for $x = f,\phi)$, and
    they give the value of $\omega$ for which the indicated
    $v$ term, given in \Eq{vterms}, is undefined, in the sense
    discussed in the text.    
  }
  \end{center}
\end{table}

The bottom line is that near the resonance ranges
indicated in the \Table{table:cases}, the integrals in \Eq{vterms}
must be handled following the principal value prescription,
and approximations such as those we have indicated in
\Eqs{pzero}{highlowkappa}, which are commonly employed,
are not valid in the cases we are considering.

Moreover, in those energy ranges, the corresponding damping term is not
negligible. This can be seen from the calculation of the imaginary part of
the self-energy, or equivalently $V^\mu_i$, in \Rref{nsnuphidamp}.
We will borrow the results of those calculations here
without further ado. But regarding $V_r$, our proposal
is to go back to \Eq{vterms} and evaluate those terms
in a systematic way that is valid through the entire neutrino energy range. 

\section{Neutrino effective potential in the $\Omega_f$ region}
\label{sec:resonance-neutrino}

For definiteness, we consider first the neutrino case in detail.
To be clear and precise, in what follows we assume
\beq
\label{positivedelta}
m_\phi > m_f\,,
\eeq
throughout. The opposite case can be treated in a similar way
by making appropriate changes.

According to \Table{table:cases},
the $v_{\bar f}$ and $v_{\bar\phi}$ terms have a resonance for
$\omega = \Omega_f$ and $\omega = \Omega_\phi$, respectively,
which we write in the form
\beqa
\label{Omegafphi}
\Omega_f & = & \frac{\Delta^2_{\phi f}}{2m_f}\,,\nonumber\\
\Omega_\phi & = & \frac{\Delta^2_{\phi f}}{2m_\phi}\,,
\eeqa
where
\beq
\label{Deltaphif}
\Delta^2_{\phi f} = m^2_\phi - m^2_f\,.
\eeq
We assume that $m_f$ and $m_\phi$ are significantly different,
such that $\Omega_{f,\phi}$ are sufficiently far apart
and the two resonance regions do not overlap. For the purpose of evaluating
the integrals this assumption is not strictly necessary,
but the physical picture is conceptually clearer if we adopt it.
Under this assumption we can consider each region separately.
Thus we consider first the region
\beq
\omega \sim \kappa \sim \Omega_f\,,
\eeq
which is the resonance region of $v_{\bar f}(\omega,\vkappa)$.

From \Eqs{Veffstandard}{Vrv} we then have
\beq
\label{VeffUr}
V^{(\nu)}_{\text{eff}}(\vkappa) = U + r\,,
\eeq
where
\beq
\label{defU}
U \equiv v_{\bar f}(\kappa,\vkappa)\,,
\eeq
and
\beq
\label{defr}
r = v_f(\kappa,\vkappa) +
v_\phi(\kappa,\vkappa) + v_{\bar\phi}(\kappa,\vkappa)\,.
\eeq
As already mentioned, the evaluation of $r$ is straightforward.
For example, let us consider the case that the $f$ background
can be treated in the non-relativistic limit. In this
case $v_f(\omega,\vkappa)$ reduces to
\beq
\label{vfcase1}
v_f(\omega,\vkappa) =
\frac{\frac{1}{8}|\lambda|^2 n_{f}}{m_f(\omega + \Omega_f)}\,.
\eeq
Since we are considering the case $m_\phi > m_f$, the $\phi$ and $\bar\phi$
backgrounds must be considered in the non-relativistic limit as well.
Thus in this case,
\begin{subequations}
\label{vphiphibarcase1exact}
\begin{align}
\label{vphicase1exact}
v_\phi(\omega,\vkappa) & = 
\frac{\frac{1}{4}|\lambda|^2 n_{\phi}}{m_\phi(\omega + \Omega_\phi)}\,,\\
\label{vphibarcase1exact}
v_{\bar\phi}(\omega,\vkappa) & = 
\frac{\frac{1}{4}|\lambda|^2 n_{\bar\phi}}{m_\phi(\omega - \Omega_\phi)}\,.
\end{align}
\end{subequations}
Since we are considering $m_\phi > m_f$ and assuming
that these masses are such that the resonance
regions $\omega \sim \Omega_f$ and $\omega \sim \Omega_\phi$ are well
separated, for the case of neutrino propagation near $\omega \sim \Omega_f$
we can put $\omega \gg \Omega_\phi$ in \Eq{vphiphibarcase1exact}.
Thus,
\beqa
\label{vphicase1}
v_\phi(\omega,\vkappa) & \simeq &
\frac{\frac{1}{4}|\lambda|^2 n_{\phi}}{m_\phi\omega}\,,
\nonumber\\
v_{\bar\phi}(\omega,\vkappa) & \simeq &
\frac{\frac{1}{4}|\lambda|^2 n_{\bar\phi}}{m_\phi\omega}\,,
\eeqa
and therefore
\beq
\label{rsol1}
r = |\lambda|^2 \left[\frac{n_{f}}{8m_f(\kappa + \Omega_f)} +
  \frac{(n_{\phi} + n_{\bar\phi})}{4m_\phi\kappa}
\right]\,.
\eeq

For later reference, it is useful to record that for $\omega$ and $\kappa$
away from the resonance region, $v_{\bar f}$ is given
by a formula analogous to those quoted above for the other
potential terms,
\beq
\label{vfbaraway}
v_{\bar f}(\omega,\vkappa) =
\frac{\frac{1}{8}|\lambda|^2 n_{\bar f}}{m_f(\omega - \Omega_f)} =
\frac{\frac{1}{8}|\lambda|^2 n_{\bar f}}{m_f}
\left\{\begin{array}{cc}
\frac{1}{\omega} & (\omega \gg \Omega_f)\\
-\frac{1}{\Omega_f} & (\omega \ll \Omega_f)\,.
\end{array}\right.
\eeq
Regarding the damping terms defined in \Eqs{defgammanu}{gammastandard}
we can borrow literally the results given in \Rref{nsnuphidamp}.

\subsection{Solution}
\label{subsec:resonance-neutrino-Omegaf}

Away from the kinematic points where $D_f(k,-p)$ does not vanish,
the procedure of taking the principal value is not necessary.
But if the kinematics is such that the integration covers
the point at which $D_f(k,-p) = 0$, the principal value operation
defines the integral around that point.

We assume the $\bar f$ gas can be treated in the non-relativistic (NR) limit.
Therefore we take
\beq
p^\mu = (m_f,\vec p)\,.
\eeq
Then doing the angular integral, remembering the
principal value prescription, we get
\beq
\label{vbarfexact0}
v_{\bar f}(\omega,\vkappa) = \frac{|\lambda|^2}{8\pi^2}
\int dp\;p^2 \frac{f_{\bar f}(p)}{2p\kappa}
\log\left|\frac{k^2 + 2m_f\omega + 2p\kappa - \Delta^2_{\phi f}}
{k^2 + 2m_f\omega - 2p\kappa - \Delta^2_{\phi f}}\right|\,,
\eeq
Further, we put
\beq
k^2 = (\omega + \kappa)(\omega - \kappa) \rightarrow 2\kappa(\omega - \kappa)\,,
\eeq
so that
\beq
k^2 + 2m_f\omega \pm 2p\kappa - \Delta^2_{\phi f} =
2\kappa\left[\left(1 + \frac{m_f}{\kappa}\right)(\omega - \kappa) \pm p +
  \frac{m_f}{\kappa}(\kappa - \Omega_f)\right]\,,
\eeq
and therefore
\beq
\label{vbarfexact}
v_{\bar f}(\omega,\vec\kappa) = \frac{|\lambda|^2}{16\pi^2\kappa}
\int dp\;p f_{\bar f}(p)
\log\left|\frac{\omega - \kappa + b + ap}{\omega - \kappa + b - ap}\right|\,,
\eeq
with
\beqa
\label{defab}
a & = & \frac{1}{1 + \frac{m_f}{\kappa}}\,,\nonumber\\
b & = & \frac{\kappa - \Omega_f}{1 + \frac{\kappa}{m_f}}\,.
\eeqa
Our next job is to evaluate the integral
\beq
\label{Iomegakappa}
I(\omega,\kappa) = \int dp\;p f_{\bar f}(p)
\log\left|\frac{\omega - \kappa + b + ap}{\omega - \kappa + b - ap}\right|\,,
\eeq
which is of the form encountered in the original calculation
by Weldon\cite{weldon-fermions} and similar calculations of the fermion
self-energy in various physical contexts\cite{fermioncalculations}.

\subsection{Evaluation of $v_{\bar f}$ for a Fermi gas}

For definiteness, we consider the case in which the $\bar f$ background
can be treated in the completely degenerate limit. We then write
\beq
\label{vbarffermi1}
v_{\bar f}(\omega,\vec\kappa) = \frac{|\lambda|^2}{16\pi^2\kappa}
I_F(\omega,\kappa)\,,
\eeq
where
\beq
\label{defIF}
I_F(\omega,\kappa) =
\int^{p_F}_0 dp\; p\log\left|\frac{p + A}{p - A}\right|\,,
\eeq
with
\beq
\label{defA}
A = \frac{\omega - \kappa + b}{a} = \omega - \kappa +
\frac{m_f}{\kappa}(\omega - \Omega_f)\,.
\eeq
A straightforward evaluation of the integral in \Eq{defIF} leads to
\beq
\label{IFformula}
I_F(\omega,\kappa) =
\frac{1}{2}(p^2_F - A^2)\log\left|\frac{p_F + A}{p_F - A}\right| + A p_F\,.
\eeq
We can now use this to find the expression for the effective
potential (in a NR Fermi gas) which, we repeat, is valid for the
entire range of the neutrino momentum.
$v_{\bar f}(\omega,\vkappa)$ is given by \Eq{vbarffermi1},
with $A$ defined in \Eq{defA}.

\subsection{Formula for $v_{\bar f}(\kappa,\vkappa)$}

First of all, as a check, let us consider the limit of small $p_F$.
Expanding the $\log$ function up to terms of order $p^3_F$,
it follows that
\beq
I_F = \frac{2p^3_F}{3A}\,.
\eeq
From \Eq{vbarffermi1}, this gives
\beq
\label{vbarfsmallpF}
v_{\bar f}(\omega,\vkappa) \simeq
\frac{\frac{1}{8}|\lambda|^2 n_{\bar f}}{m_f(\omega - \Omega_f)}\,,
\eeq
where we have used $p^3_F = 3\pi^2 n_{\bar f}$, and from \Eq{defA},
$A \sim \frac{m_f}{\kappa}(\omega - \Omega_f)$ for $\omega \sim \kappa$.
It is reassuring to see that the formula
for $v_{\bar f}(\omega,\vkappa)$ in \Eq{vbarfsmallpF}
coincides with \Eq{vfbaraway},
which is obtained by taking $p^\mu\rightarrow (m_f,\vec 0)$ from the beginning
in the integrand. However, as we have emphasized,
this limiting form is not valid for values of $\kappa$
near the resonance point.

\begin{figure}
\begin{center}
\epsfig{file=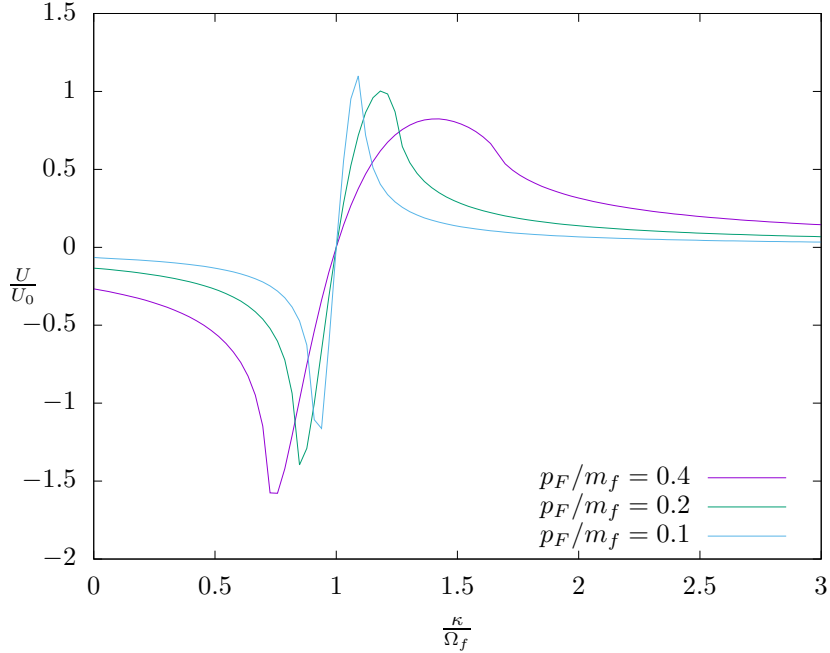,bbllx=168,bblly=272,bburx=476,bbury=517}
\end{center}
\caption[]{
  Plot of $U(\kappa)$, for $\kappa \sim \Omega_f$,
  with some example values of $p_F/m_f$.
  \label{fig:v}
}
\end{figure}

In the general case, going back to \Eq{IFformula},
\beq
\label{IFformulakk}
\frac{I_F(\kappa,\kappa)}{p^2_F} = \frac{1}{2}(1 - \eta^2)
\log\left|\frac{1 + \eta}{1 - \eta}\right| + \eta\,,
\eeq
where
\beq
\label{eta}
\eta = \frac{m_f}{p_F}\frac{\xi - 1}{\xi}\,,
\eeq
with
\beq
\label{xi}
\xi = \frac{\kappa}{\Omega_f}\,.
\eeq
Using \Eq{IFformulakk} in \Eq{vbarffermi1},
\beq
\label{vfbarfinal}
U(\kappa) = v_{\bar f}(\kappa,\vkappa) = U_0 \frac{1}{\xi}
\left[\frac{1}{2}(1 - \eta^2)
  \log\left|\frac{1 + \eta}{1 - \eta}\right| + \eta\right]\,,
\eeq
where
\beq
U_0 = \frac{|\lambda|^2 p^2_F}{16\pi^2\Omega_f}\,.
\eeq
The formula in \Eq{vfbarfinal} is our main result.
For reference, a plot of $U(\kappa)$ is shown in \Fig{fig:v}.

We note the following. Away from the resonance region, $\eta$
has the limiting values
\beq
\label{bnonresonance}
\eta = \frac{m_f}{p_F}\left\{
\begin{array}{ll}
  -\frac{1}{\xi} & \mbox{(low $\kappa$)}\\
   1 & \mbox{(high $\kappa$)}\,.
\end{array}
\right.
\eeq
Therefore, either in the high ($\kappa \gg \Omega_f$)
or low ($\kappa \ll \Omega_f$) momentum limit, $\eta$
is large and we can approximate $I_F(\kappa,\vkappa)$ by its
limiting value for large $\eta$, which gives
\beq
\label{IFlargeeta}
\frac{I_F(\kappa,\vkappa)}{p^2_F} \simeq \frac{2}{3\eta} =
\frac{2p_F}{3m_f}\left(\frac{\xi}{\xi - 1}\right)\,.
\eeq
Then from \Eq{vbarffermi1}, this reproduces, again, \Eq{vbarfsmallpF}.
Thus, in these asymptotic limits, namely high or low momentum
as specified above, the expression for
$v_{\bar f}(\kappa,\vkappa)$ given in \Eq{vfbarfinal} coincides with the results
that are obtained by approximating from the beginning
the integrals for $v_{\bar f}$ under the same conditions, namely,
away from the resonance and for the non-relativistic limit.
However, as we can see from \Eq{eta}, $|\eta| < 1$ for values of
$\kappa$ such that $|\xi - 1| < p_F/m_f$, so that in this range
the asymptotic form of $I_F$ given in \Eq{IFlargeeta} is not valid.

The virtue of \Eq{vfbarfinal} is that it is valid also
in the resonance region, interpolating between the asymptotic
expressions corresponding to high or low $\kappa$ mentioned above,
and they allow us to consider the propagation in the resonance
region, including the range $|\kappa/\Omega_f - 1| < p_F/m_f$ mentioned.
On the other hand, the imaginary part of the dispersion relation
is important in that region, as we have remarked, and it must be
included in the treatment of the propagation equation.
As a guide to possible applications we consider that next.

\section{Damping term}
\label{sec:damping-Omegaf}

The damping term $\gamma^{(\nu)}$ in \Eq{defgammanu}
is not negligible in the resonance region.
As indicated in \Eq{gammastandard}, it is determined
from the absorptive part $V^\mu_i$ of the effective potential which in turn
is determined from the $\Sigma_{12}$ element of the thermal
self-energy matrix.
That calculation was carried out in detail
in \Rref{nsnuphidamp}. As shown in that reference, the resulting formula
for $\gamma^{(\nu)}$ is related
to the transition probabilities for the various processes in
which the neutrino may be annihilated or created,
such as $\nu + \bar f \leftrightarrow \bar\phi$ and
$\nu + \phi \leftrightarrow f$, in the forward and reverse directions.
The formulas involve the phase space integrals
weighted by the appropriate momentum distribution functions.
Here we just need to borrow the results of those calculations.
Quoting the results that are summarized in Eq.\ (3.38) of \Rref{nsnuphidamp},
\beq
\label{gammanuisotropicresult}
\frac{\gamma^{(\nu)}(\kappa)}{2} = \frac{|\lambda|^2}{32\pi\kappa^2\beta}
\left\{
\begin{array}{ll}
\Delta^2_{f\phi}\left[
\log\left(1 + e^{-\beta E^{(I)}_{min} + \alpha_f}\right) -
\log\left(1 - e^{-\beta \Omega^{(I)}_{min} + \alpha_\phi}\right)
\right] & (m_f > m_\phi) \\[12pt]
\Delta^2_{\phi f}\left[
\log\left(1 + e^{-\beta E^{(II)}_{min} - \alpha_f}\right) -
\log\left(1 - e^{-\beta \Omega^{(II)}_{min} - \alpha_\phi}\right)
\right] & (m_\phi > m_f)\,.
\end{array}\right.
\eeq
The corresponding formulas for
$\frac{\gamma^{(\bar\nu)}(\kappa)}{2}$ are obtained by making the
substitutions
\beq
\label{gammanubisotropicresult}
\alpha_{f,\phi} \rightarrow - \alpha_{f,\phi}\,.
\eeq

To be consistent with \Eq{positivedelta}, we focus on the second
formula in \Eq{gammanuisotropicresult}. In that formula,
\beqa
\label{Ebarxmin}
\Omega^{(II)}_{min} & = & \frac{\Delta^2_{\phi f}}{4\kappa} +
\frac{\kappa m^2_\phi}{\Delta^2_{\phi f}}\,,\nonumber\\
E^{(II)}_{min} & = & \frac{\Delta^2_{\phi f}}{4\kappa} +
\frac{\kappa m^2_f}{\Delta^2_{\phi f}}\,,
\eeqa
with $\Delta^2_{\phi f}$ defined in \Eq{Deltaphif}.
The term involving $\alpha_f$ corresponds to the
contribution from the $\bar f$ gas in the background,
while the term with $\alpha_\phi$ corresponds to the $\bar\phi$ gas,
which are associated with the processes
$\nu + \bar f\leftrightarrow \bar\phi$ and
$\nu + \phi\leftrightarrow f$, respectively.

To complement our calculation of the effective potential
in the previous section, here we want to calculate the
$\bar f$ background contribution to the damping in the case
that it can be considered as a completely degenerate fermion gas.
In order to bring out the physical picture in a clearer
way, let us consider first the case that both the $\bar\phi$ and $\bar f$
gases can be treated in the classical and non-relativistic limit.

\subsection{Damping in the classical and non-relativistic (NR) limit}

In that case
\beqa
\frac{\gamma^{(\nu)}(\kappa)}{2} & = &\frac{|\lambda|^2}{32\pi\kappa^2\beta}
\Delta^2_{\phi f}\left[e^{-\beta E^{(II)}_{min} - \alpha_f} +
e^{-\beta \Omega^{(II)}_{min} - \alpha_\phi}\right]\,,
\eeqa
where in the non-relativistic limit (as we have assumed in
\Section{subsec:resonance-neutrino-Omegaf}), the chemical
potentials are\footnote{%
  These are obtained by requiring
  \begin{displaymath}
    n_{\bar x} = g_x\int\frac{d^3p}{(2\pi)^3}e^{-\beta E_x - \alpha_x}
    \qquad (x =f,\phi)\,,
  \end{displaymath}
  with $g_f = 2g_\phi = 2$.
}
\beqa
e^{-\alpha_f} & = & \frac{1}{2}n_{\bar f}
\left(\frac{2\pi\beta}{m_f}\right)^\frac{3}{2}e^{\beta m_f}\,,\nonumber\\
e^{-\alpha_\phi} & = & n_{\bar\phi}
\left(\frac{2\pi\beta}{m_\phi}\right)^\frac{3}{2}e^{\beta m_\phi}\,.
\eeqa
Therefore,
\beq
\label{dampingnuNRclassical}
\frac{\gamma^{(\nu)}(\kappa)}{2} =
\frac{\Omega^2_f}{\kappa^2}\frac{\gamma^{(0)}_{\bar f}}{2}
e^{-\Lambda_f} +
\frac{\Omega^2_\phi}{\kappa^2}
\frac{\gamma^{(0)}_{\bar\phi}}{2} e^{-\Lambda_\phi}\,,
\eeq
where
\beqa
\label{gammanu0Lambdas}
\frac{\gamma^{(0)}_{\bar f}}{2} & = & 
\frac{|\lambda|^2}{16}\sqrt{\frac{2\pi T}{m_f}}\frac{n_{\bar f}}{T\Omega_f}
\,,\nonumber\\
\frac{\gamma^{(0)}_{\bar \phi}}{2} & = & 
\frac{|\lambda|^2}{8}\sqrt{\frac{2\pi T}{m_\phi}}
\frac{n_{\bar\phi}}{T\Omega_\phi}
\,,\nonumber\\
\Lambda_f & = & \frac{m_f}{2T}
\frac{\left(\frac{\kappa}{\Omega_f} - 1\right)^2}
{\left(\frac{\kappa}{\Omega_f}\right)}\,,\nonumber\\
\Lambda_\phi & = & \frac{m_\phi}{2T}
\frac{\left(\frac{\kappa}{\Omega_\phi} - 1\right)^2}
{\left(\frac{\kappa}{\Omega_\phi}\right)}\,.
\eeqa
with $\Omega_{f,\phi}$ defined in \Eq{Omegafphi}.

The picture that emerges is consistent with our previous
discussions regarding the resonances and complements it in a
practical way. Outside of either resonance range the damping
is exponentially small and can be neglected in the formula for the dispersion
relation. Therefore, if we are considering a neutrino propagating in the
$\Omega_f$ resonance region, we can discard the $\bar\phi$ contribution
to the damping term, assuming, as we do, that $m_f$ and $m_\phi$
are sufficiently different that the resonances at $\Omega_f$ and $\Omega_\phi$
do not overlap.

\begin{figure}
\begin{center}
\epsfig{file=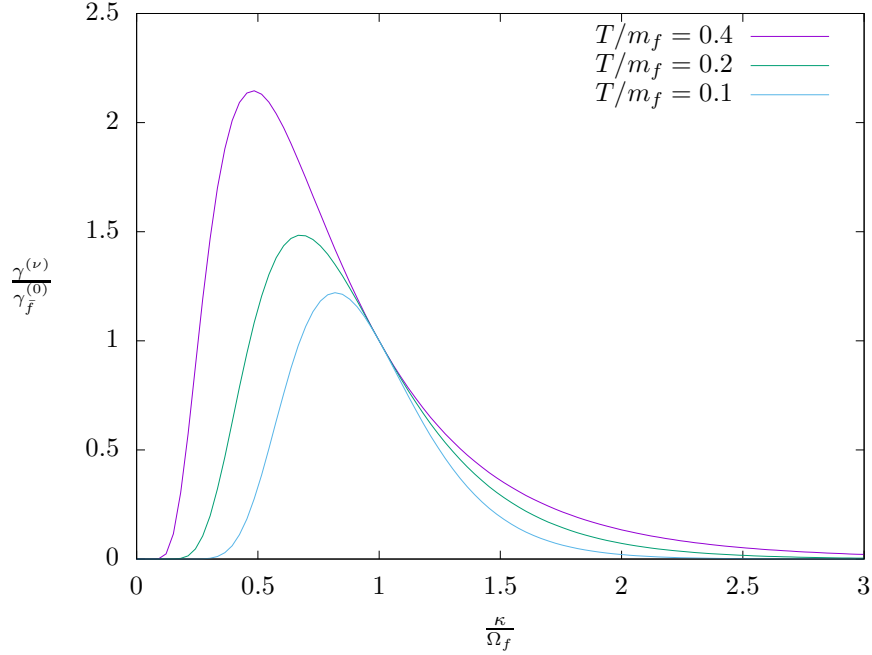,bbllx=153,bblly=272,bburx=476,bbury=517}
\end{center}
\caption[]{
  Plot of the damping term for $\kappa \sim \Omega_f$,
  given by \Eq{dampingnuNRclassicalfregion}, for
  some example values of $T/m_f$.
  \label{fig:gamma-mb}
}
\end{figure}

For illustrative purposes we show a plot of $\gamma^{(\nu)}(\kappa)$
in \Fig{fig:gamma-mb} in the $\kappa \sim \Omega_f$ resonance region,
obtained as follows. As already stated,
we assume that $m_f$ and $m_\phi$ are sufficiently
different that the resonances at $\Omega_f$ and $\Omega_\phi$ do not overlap;
e.g., the $\kappa = \Omega_\phi$ point falls outside
the range shown in the plot.
Under this condition, in the $\Omega_f$ resonance region
we can neglect the $\bar\phi$ contribution in \Eq{dampingnuNRclassical}
and consider simply
\beq
\label{dampingnuNRclassicalfregion}
\frac{\gamma^{(\nu)}}{\gamma^{(0)}_{\bar f}} =
  \frac{1}{\xi^2} e^{-\frac{m_f}{2T}\frac{(\xi - 1)^2}{\xi}}\,,
\eeq
with $\xi$ defined in \Eq{xi}.
Equation (\ref{dampingnuNRclassicalfregion}) is plotted in \Fig{fig:gamma-mb}.

\subsection{Damping in the degenerate limit}

Armed with the results of the previous section, we thus ignore
the $\bar\phi$ contribution to the damping, and as a complement
to the formula for the effective potential in \Eq{vfbarfinal} here
we calculate the $\bar f$ background contribution
to $\gamma^{(\nu)}$, that is
\beq
\label{defgammadeg}
\frac{\gamma^{(\nu)}(\kappa)}{2} =
\frac{|\lambda|^2\Delta^2_{\phi f}}{32\pi\kappa^2\beta}
\log\left(1 + e^{-\beta E^{(II)}_{min} - \alpha_f}\right)\,,
\eeq
in the limit of a completely degenerate fermion gas.
This formula, and the results discussed below, hold in all
the kinematic regime of the fermion gas, so they can be used
in the NR case, or in any other case as well.

Setting
\beq
\alpha_f = -\beta E_F\,,
\eeq
where $E_F$ is the Fermi energy, and taking the degenerate limit
($\beta \rightarrow \infty$),
\beq
\label{gammanuex1}
\frac{\gamma^{(\nu)}(\kappa)}{2} =
\frac{|\lambda|^2\Delta^2_{\phi f}}{32\pi\kappa^2}
\left(E_F - E^{(II)}_{min}\right)
\theta\left(E_F - E^{(II)}_{min}\right)\,.
\eeq
The step function in \Eq{gammanuex1} implies that $\gamma^{(\nu)}(\kappa)$
is non-zero if $\kappa$ lies in the range such that
\beq
\label{kappalimitex1}
\frac{1}{2}(E_F - p_F) \leq \frac{\kappa m^2_f}{\Delta^2_{\phi f}} \leq
\frac{1}{2}(E_F + p_F)\,,
\eeq
or it is zero otherwise\footnote{%
To prove \Eq{kappalimitex1} we
rewrite the condition $E_F \geq E^{(II)}_{min}$ in the form
\beq
\label{conditionforx}
\frac{1}{4x} + x \leq \frac{E_F}{m_f}\,,
\eeq
where
\beq
x \equiv \frac{\kappa m_f}{\Delta^2_{\phi f}}\,.
\eeq
The left-hand-side of \Eq{conditionforx} can be written in the form
\beq
\label{conditionforxlhs}
\frac{1}{4x} + x = \frac{1}{x}(x - x_{+}(t))(x - x_{-}(t)) + t
\eeq
where $x_{\pm}(t)$, given by
\beq
x_{\pm}(t) \equiv \frac{t}{2} \pm \frac{1}{2}\sqrt{t^2 - 1}\,,
\eeq
satisfy
\beq
\label{eqforx}
\frac{1}{4x_{\pm}} + x_{\pm} = t\,,
\eeq
for any $t \geq 1$.
The functions $x_{\pm}(t)$ are positive, have the same value at $t = 1$, and as
$t$ increases $x_{+}$ increases while $x_{-}$ decreases. Therefore
for any value of $x$ such that
\beq
x_{-}(t) \leq x \leq x_{+}(t)\,,
\eeq
$\frac{1}{x}(x - x_{+}(t))(x - x_{-}(t)) \leq 0$, and therefore
from \Eq{conditionforxlhs}
\beq
\frac{1}{4x} + x < t\,,
\eeq
for any such value of $x$.
It then follows that all the values of $x$ that lie between $x_{-}(E_F/m_f)$
and $x_{+}(E_F/m_f)$ satisfy \Eq{conditionforx}, while the values outside
that range will violate it. Using the fact that
\beq
x_{\pm}(E_F/m_f) = \frac{1}{2}\left(\frac{E_F}{m_f}\right)
\pm \frac{1}{2}\sqrt{\left(\frac{E_F}{m_f}\right)^2 - 1} =
\frac{1}{2}\left(\frac{E_F}{m_f}\right)
\pm \frac{1}{2}\frac{p_F}{m_f}\,,
\eeq
proves \Eq{kappalimitex1}.}.
Equation (\ref{gammanuex1}) can be written in the form
\beq
\label{gammanufermi}
\gamma^{(\nu)} = \gamma^{(0)}_{f}
\frac{1}{\xi^2}\left[
  \frac{E_F}{m_f} - \frac{1}{2}\left(\xi + \frac{1}{\xi}\right)
  \right]\,,
\eeq
for $\xi$, defined in \Eq{xi}, in the range
\beq
\frac{1}{m_f}(E_f - p_F) \leq \xi \leq \frac{1}{m_f}(E_F + p_F)\,,
\eeq
and
\beq
\frac{\gamma^{(0)}_{f}}{2} = 
\frac{|\lambda|^2 m^2_f}{16\pi\Omega_f}\,.
\eeq
As already stated, the result given in \Eq{gammanufermi} holds in all
the kinematic regime of the fermion gas.

\begin{figure}
\begin{center}
\epsfig{file=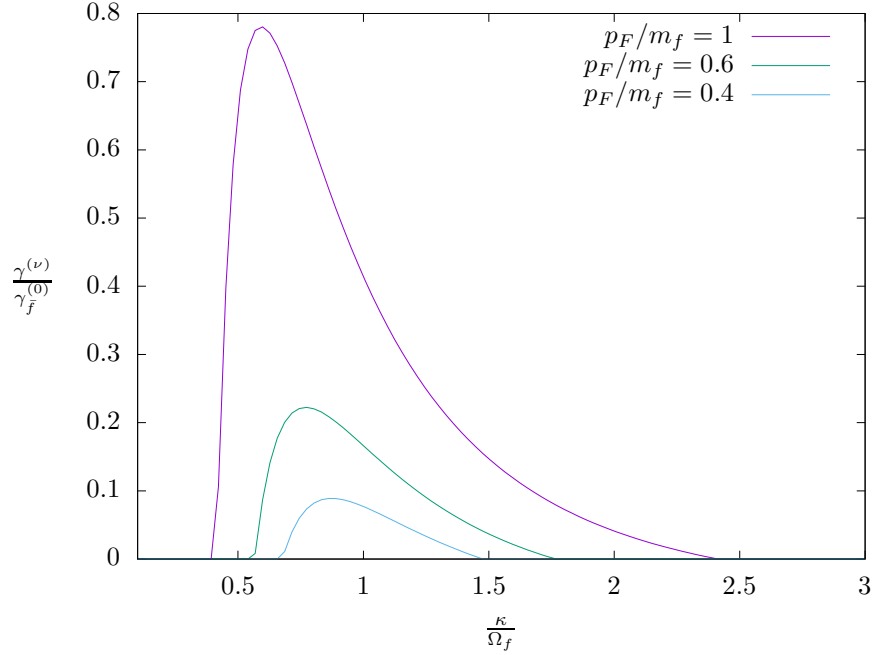,bbllx=153,bblly=272,bburx=476,bbury=517}
\end{center}
\caption[]{
  Plot of the damping term in the case of a degenerate
  $\bar f$ background, given by \Eq{gammanufermi},
  for some example values of $p_F/m_f$.
  \label{fig:gamma-fermi}
}
\end{figure}

\Eq{gammanufermi} is ploted in \Fig{fig:gamma-fermi}. The damping
becomes smaller as $p_F$ decreases.  From \Fig{fig:v} we see that,
at the same time, the width of the resonance peaks in the effective potential
become narrower. These features indicate that the resonance effects
are more important for relatively large values of $p_F$ (high density)
and less important as $p_F$ decreases (lower density). 
A similar effect occurs with the damping in the classical case,
which reduces as the temperature decreases, as shown in \Fig{fig:gamma-mb}.

\section{Discussion}
\label{sec:discussion}

Here we discuss some generalizations and extensions of our work. On
one hand, the case of antineutrinos, as well as the other resonance region,
around $\omega \sim \Omega_\phi$ in the notation of \Table{table:cases},
can be treated in analogous form. In order to point out possible
differences in details and/or implementation we consider them briefly here.

On the other hand, since we have restricted ourselves to the case of one
neutrino flavor propagating and interacting in the $f$ and $\phi$ background,
as a guide and example to possible applications and generalizations,
here we will consider the application to the oscillation equations
including the damping term, in a simple two-generation case.

\subsection{Anti-neutrino propagation near $\Omega_f$}
\label{subsec:resonance-antineutrino-Omegaf}

Again we assume that $m_\phi > m_f$, and we consider the propagation
at energies $\omega^{(\bar\nu)} \sim \Omega_f$. This is the resonance
region of the term $v_f$, which is the term that must be singled
out for special consideration.
From \Eqs{Veffstandard}{Vrv} we have in the present case
\beq
V^{(\bar\nu)}_{\text{eff}}(\vkappa) = \bar U + \bar r\,,
\eeq
where
\beq
\label{defUbar}
\bar U = -v_{f}(-\kappa,-\vkappa) \,,
\eeq
and
\beq
\label{defrbar}
\bar r = -v_{\bar f}(-\kappa,-\vkappa) -
v_\phi(-\kappa,-\vkappa) - v_{\bar\phi}(-\kappa,-\vkappa)\,.
\eeq
For $v_\phi$ and $v_{\bar\phi}$, we can simply borrow the formulas
given in \Eqs{vphicase1exact}{vphicase1}, while for
$v_{\bar f}(\omega,\vkappa)$ the corresponding formula
in this case is \Eq{vfbaraway}.
Thus, mimicking the steps leading to \Eq{rsol1}, we obtain
\beq
\label{rbsol1}
\bar r = |\lambda|^2 \left[\frac{n_{\bar f}}{8m_f(\kappa + \Omega_f)} +
  \frac{(n_{\phi} + n_{\bar\phi})}{4m_\phi\kappa}\right]\,.
\eeq

Regarding $v_f$, we go back to \Eq{vf}. Carrying out the angular integral,
remembering the principal value prescription,
\beq
\label{vfexact0}
v_{f}(\omega,\vkappa) = -\frac{|\lambda|^2}{8\pi^2}
\int dp\;p^2 \frac{f_{f}(p)}{2p\kappa}
\log\left|\frac{k^2 - 2m_f\omega + 2p\kappa - \Delta^2_{\phi f}}
{k^2 - 2m_f\omega - 2p\kappa - \Delta^2_{\phi f}}\right|\,.
\eeq
As we can see,  $-v_{f}(-\omega,-\vkappa)$ is given
by the same expression given in \Eq{vbarfexact0}
for $v_{\bar f}(\omega,\vkappa)$, with the substitution
$f_{\bar f}\rightarrow f_f$ in the integrand. Thus,
for example, in the case that $f$ gas can be treated in the
NR and degenerate limit, the net result of this is that the final formula
for $\bar U$ is the same as the formula for $U$ given in \Eq{vfbarfinal},
but of course with $p_F$ given in terms of the $f$ number density,
$p^3_F = 3\pi^2 n_f$.

The damping can be treated similarly to the case of neutrinos
in \Section{sec:damping-Omegaf}, but in the present case the relevant
number densities are $n_{f}$ and $n_{\phi}$.
Explicitly, remembering \Eq{gammanubisotropicresult},
in the classical and NR limit the damping is
\beq
\label{dampingnubNRclassical}
\frac{\gamma^{(\bar\nu)}(\kappa)}{2} =
\frac{\Omega^2_f}{\kappa^2}\frac{\gamma^{(0)}_{f}}{2}
e^{-\Lambda_f} +
\frac{\Omega^2_\phi}{\kappa^2}
\frac{\gamma^{(0)}_{\phi}}{2} e^{-\Lambda_\phi}\,,
\eeq
where
\beqa
\label{gammanu0bLambdas}
\frac{\gamma^{(0)}_{f}}{2} & = & 
\frac{|\lambda|^2}{16}\sqrt{\frac{2\pi T}{m_f}}\frac{n_{f}}{T\Omega_f}\,,
\nonumber\\
\frac{\gamma^{(0)}_{\phi}}{2} & = & 
\frac{|\lambda|^2}{8}
\sqrt{\frac{2\pi T}{m_\phi}}\frac{n_{\phi}}{T\Omega_\phi}\,,
\eeqa
with $\Lambda_{f,\phi}$ given by \Eq{gammanu0Lambdas}.
For a degenerate $f$ gas, the formula for $\gamma^{(\bar\nu)}$
is the same as \Eq{gammanufermi}, but with the reinterpretation
of $p_F$ as the Fermi momentum associated with the $f$ gas number density
$n_f$, as we have stated above. The sketches of the damping and the
effective potential in this case are therefore similar to those shown in
\Figss{fig:v}{fig:gamma-mb}{fig:gamma-fermi}
with the corresponding identification of the parameters involved.

In the case of a neutrino or antineutrino propagating near the
$\Omega_\phi$ energy region, the same method can be applied to calculate
the effective potential.
In this case the term that must be singled out for special treatment
is $v_{\bar\phi}$ (or $v_{\phi}$ for antineutrinos). The relevant integrals
are of the same form given in \Eq{Iomegakappa}, but they
involve the $\bar\phi$ or $\phi$ distribution functions. In the classical
limit of the distribution functions, the task involves the computation
of the generic integrals
\beq
\int dx\, e^{-x^n}x\log\left|\frac{x + A}{x - A}\right|\,,
\eeq
with $n = 1,2$ in the ultra-relativistic and non-relativistic limits,
respectively. Integrals of this form appear in similar calculations
in other contexts as already mentioned\cite{fermioncalculations}.
We do not pursue this case any further here.

\subsection{Two-generation example}
\label{subsec:twogenexample}

We consider a two-generation case, assuming that only the
first generation (e.g., electron neutrino) couples to $f$ and $\phi$.
Working at the level of the evolution of the flavor spinor,
the equation, including the damping term, is
\beq
i\partial_t\phi = \left(H_r - \frac{i}{2}\Gamma\right)\phi\,.
\eeq

Up to a term proportional to identity matrix, the Hamiltonian is
\beq
H_{r} = \frac{1}{2}\frac{\Delta m^2_{21}}{2\kappa}
  \left(\begin{array}{cc}
    -\cos 2\theta & \sin 2\theta\\ \sin 2\theta & \cos 2\theta
\end{array}\right) +
  \frac{1}{2}\left(\begin{array}{cc}
  V^{(\nu)}_{\text{eff}} & 0 \\ 0 & -V^{(\nu)}_{\text{eff}}
\end{array}\right)\,,
\eeq
where $\Delta m^2_{21} = m^2_2 - m^2_1$, while
\beq
\Gamma =
\left(\begin{array}{cc}
  \gamma^{(\nu)} & 0 \\ 0 & 0
\end{array}
\right) \equiv \gamma^{(\nu)} I_e\,,
\eeq
with
\beq
I_e \equiv
\left(\begin{array}{cc}
  1 & 0 \\ 0 & 0
\end{array}
\right)\,.
\eeq
$V^{(\nu)}_{\text{eff}}$ and $\gamma^{(\nu)}$ are understood to be the
effective potential and damping term that we have obtained.

Following the usual steps, $H_r$ can be written in the standard form
\beq
\label{Hrstd}
H_r = \frac{h}{2}\left(\begin{array}{cc}
    -\cos 2\theta_m & \sin 2\theta_m\\
    \sin 2\theta_m & \cos 2\theta_m
\end{array}\right)\,,
\eeq
where
\beq
\label{hmag}
h = \frac{\Delta^2_m}{2\kappa}\,,
\eeq
and
\beqa
\label{cos2thetamdef}
\label{cos2thetam}
\label{sin2thetam}
\cos 2\theta_m & = &
\frac{1}{\Delta^2_m}
\left(\Delta m^2_{21}\cos 2\theta - 2\kappa V^{(\nu)}_{\text{eff}}\right)\,,
\nonumber\\
\sin 2\theta_m & = &
\frac{\Delta m^2_{21}\sin 2\theta}{\Delta^2_m}\,,
\eeqa
with
\beq
\label{Deltam}
\Delta^2_m \equiv \sqrt{(\Delta m^2_{21}\sin 2\theta)^2 +
\left(\Delta m^2_{21}\cos 2\theta - 2\kappa V^{(\nu)}_{\text{eff}}\right)^2}\,.
\eeq

Writing the solution in the form
\beq
\phi(t) = G\phi(0)\,,
\eeq
in the absence of the damping term
\beq
\label{evolutionU}
G = \sum_{s = \pm} u_s u^\dagger_s e^{-i\lambda_s t}\,,
\eeq
where $\lambda_s = sh$ are the eigenvalues of $H_r$,
and $u_s$ the corresponding eigenspinors
\beqa
u_{+} & = & \left(\begin{array}{c}
  \sin\theta_m \\ \cos\theta_m
\end{array}\right)\,,\nonumber\\[12pt]
u_{-} & = & \left(\begin{array}{c}
  \cos\theta_m \\ -\sin\theta_m
\end{array}\right)\,.
\eeqa

In the spirit of a perturbative treatment
of the damping term, we construct the solution by taking the
eigenvectors of $H$ to be the same as the eigenvectors of $H_r$,
but with the eigenvalues modified by the first order corrections; i.e.,
\beq
\lambda_{s} = sh - \frac{i}{2}u^\dagger_s\Gamma u_s\,.
\eeq
By explicitly calculation,
\beq
\lambda_{s} = \left\{\begin{array}{cc}
  h - \frac{i}{2}\gamma_s & (s = +)\\
  -h - \frac{i}{2}\gamma_c & (s = -)\,,
\end{array}\right.
\eeq
where we have defined
\beqa
\gamma_s & = & \gamma^{(\nu)}\sin^2 \theta_m\,,\nonumber\\
\gamma_c & = & \gamma^{(\nu)}\cos^2 \theta_m\,.
\eeqa

We then obtain the following explicit expression for the evolution matrix
\beq
G = u_{+} u^\dagger_{+} e^{-iht}e^{-\frac{1}{2}\gamma_s t} +
u_{-} u^\dagger_{-} e^{iht}e^{-\frac{1}{2}\gamma_c t}\,.
\eeq
So, for example, for
\beq
\phi(0) = \left(\begin{array}{c} 1 \\ 0\end{array}\right)\,,
\eeq
we have the persistence and transition amplitudes,
\beqa
\phi_1(t) & =  & \sin^2\theta_m e^{-iht}e^{-\frac{1}{2}\gamma_s t}
+ \cos^2\theta_m e^{iht}e^{-\frac{1}{2}\gamma_c t}\,,\nonumber\\
\phi_2(t) & =  & \sin\theta_m \cos\theta_m \left(
e^{-iht}e^{-\frac{1}{2}\gamma_s t} -
e^{iht}e^{-\frac{1}{2}\gamma_c t}\right)\,,
\eeqa
and the corresponding oscillation probabilities can be written in the
form\footnote{%
We have used
\begin{eqnarray*}
  \sin^4\theta_m & = & \frac{1}{4}(1 - \cos 2\theta_m)^2 =
  \frac{1}{4}(2 - 2\cos 2\theta_m - \sin^2 2\theta_m)\nonumber\\
  \cos^4\theta_m & = & \frac{1}{4}(1 + \cos 2\theta_m)^2 =
  \frac{1}{4}(2 + 2\cos 2\theta_m - \sin^2 2\theta_m)
\end{eqnarray*}
}
\beqa
\label{Poscprobs}
P_1 \equiv |\phi_1|^2 & = &
\frac{1}{2}\left(e^{-\gamma_c t} + e^{-\gamma_s t}\right) +
\frac{1}{2}\cos 2\theta_m \left(e^{-\gamma_c t} - e^{-\gamma_s t}\right)
\nonumber\\
&&\mbox{} - \frac{1}{2}\sin^2 2\theta_m\left[
  \frac{1}{2}\left(e^{-\gamma_c t} + e^{-\gamma_s t}\right) -
  e^{-\frac{1}{2}\gamma^{(\nu)} t}\cos 2ht\right]\,,\nonumber\\
P_2 \equiv |\phi_2|^2 & = & \frac{1}{2}\sin^2 2\theta_m\left[
  \frac{1}{2}\left(e^{-\gamma_c t} + e^{-\gamma_s t}\right) -
  e^{-\frac{1}{2}\gamma^{(\nu)} t}\cos 2ht\right]\,.
\eeqa

The importance of the damping terms depends on the interplay between
$h$, defined in \Eq{hmag}, and $\gamma^{(\nu)}$.
To be more specific, we can consider, for example, the propagation
of neutrinos near the $\Omega_f$ region, with $V^{(\nu)}_{\text{eff}}$
given by \Eq{VeffUr} (with $r$ and $U$ given by \Eqs{rsol1}{vfbarfinal},
respectively) and $\gamma^{(\nu)}$ by \Eq{dampingnuNRclassicalfregion}.
For small values of $\gamma^{(\nu)}$,
many oscillation length cycles are required for the damping effects to be
observable. A distinctive feature of \Eq{Poscprobs} is the energy dependence
of the damping and the oscillation terms, based on the formulas
for $\gamma^{(\nu)}$ and $V^{(\nu)}_{\text{eff}}$ that we have given.
For example, if we consider a pure $\bar f$ background (no
$f$ and no $\phi,\bar\phi$ backgrounds) then $V^{(\nu)}_{\text{eff}}$
is given only by $U$ [\Eqs{VeffUr}{vfbarfinal}], which becomes
zero for neutrino energies very close to the resonance point $\Omega_f$.
In that regime $h$ reduces to its vacuum value $\Delta m^2_{21}/2\kappa$
while the damping term reaches its largest value.

\section{Conclusions and outlook}
\label{sec:conclusions}

In this work we have been concerned with the propagation
of a neutrino in a background of fermions ($f$) and scalars ($\phi$),
interacting via a Yukawa-type interaction. Our particular goal was to
obtain the dispersion relation in the case that the neutrino
energy lies in the range in which the absorption and production processes
become kinematically accessible, such as
$\nu + \bar f \rightarrow \bar\phi$ or the crossed counterparts,
and the corresponding ones for the antineutrino.
The relevant energy values are around
$|m^2_\phi - m^2_f|/2m_f$ or $|m^2_\phi - m^2_f|/2m_\phi$,
to which we refer as a resonance energy range.
The distinguishing aspect of these energy ranges is that
the one-loop formula for the neutrino self-energy
has a singularity, which is the indication that the self-energy
acquires an imaginary part that cannot be ignored. Technically,
the imaginary part is associated with the damping effects, while
the integral formula for the real part must be
evaluated using the principal value of the integral.

Writing the dispersion relation in the form
$\omega = \kappa + V_{\text{eff}} - i\gamma/2$, we gave
the explicit formulas for the effective potential ($V_{\text{eff}}$)
and damping ($\gamma$) for some cases
that allowed us to give analytic results. In particular we considered in detail
the evaluation of those quantities for a neutrino propagating
with the energy near ($m^2_\phi - m^2_f)/2m_f$
(corresponding, for $m_\phi > m_f$, to the processes
$\nu + \bar f \rightarrow \bar\phi$ becoming kinematically accessible),
in the case that the $\bar f$ background can be treated as a non-relativistic
degenerate Fermi gas. We also considered the analogous case for
an antineutrino.

The formulas obtained have the property that,
when the neutrino energy is either much larger or much smaller
than the resonance energy, $V^{(\nu)}_{\text{eff}}$ reduces to the
effective potential that has been already determined in the literature
in the high or low momentum regime, respectively. 
The virtue of the formula we give for $V^{(\nu)}_{\text{eff}}$ is that it is
valid also in the \emph{resonance energy range}, which is outside
the two limits mentioned. We outlined how the same
strategy can be applied to consider the case of a neutrino propagating
with an energy in the other resonance region
$|m^2_\phi - m^2_f|/2m_\phi$, in which case the terms in
the effective potential that require special consideration are those
corresponding to the contribution from the scalar backgrounds.
For example, for $m_f > m_\phi$, the resonance shows up in the
$v_\phi$ contribution, which corresponds to the processes
$\nu + \phi \rightarrow f$ becoming kinematically accessible.

For definiteness we restricted ourselves to the calculation
of the dispersion relation in the case that only one neutrino flavor
interacts with the $f$ and $\phi$ background particles.
As a guide and example to possible applications and generalizations,
we gave the relevant formulas for the $V_{\text{eff}}$ and $\gamma$ matrices,
and considered the solution to the oscillation equations
including the damping term, in a simple two-generation case.

The same strategy we have used to determine the effective potential
for a neutrino propagating in an $f$ background in the
energy range to produce a $\phi$ particle, can be applied
to the case of a neutrino propagating in an electron background with an
energy in the Glashow resonance region\cite{glashowresonance}.
Several technical aspects of the calculations are of course different,
but the idea of determining the  effective potential for such
energy range by the method we have followed can be applied
to that case as well.

%\begin{acknowledgments}
The work of S. S. is partially supported by 
DGAPA-UNAM (Mexico) PAPIIT project No. IN103522.
%\end{acknowledgments}

%\bibliographystyle{ieeetr}
%\bibliography{main}

\end{document}